\documentstyle[prl,aps,epsf,epsfig,twocolumn,floats]{revtex}

\begin{document}
\newcommand{\DAVG}{$D^*$}
\newcommand{\KK}{k}
\newcommand{\PC}{\%}

\title{Quantification of depth of anesthesia by nonlinear time series analysis
of brain electrical activity}

\author{G. Widman$^a$, T. Schreiber$^b$, B. Rehberg$^c$, A. Hoeft$^c$, 
and C.~E. Elger$^a$}

\address{$^a$Clinic of Epileptology, University of Bonn,
Sigmund-Freud-Str. 25, 53105 Bonn, Germany\\
$^b$Max Planck Institute for the Physics of Complex Systems,
N\"othnitzer Strasse 38, 01187 Dresden, Germany\\
$^c$Department of Anesthesiology, University of Bonn,
Sigmund-Freud-Str. 25, 53105 Bonn, Germany}
\maketitle

\begin{abstract}
We investigate several quantifiers of the electroencephalogram (EEG) signal
with respect to their ability to indicate depth of anesthesia.  For 17 patients
anesthetized with Sevoflurane, three established measures (two spectral and one
based on the bispectrum), as well as a phase space based nonlinear correlation
index were computed from consecutive EEG epochs.  In absence of an independent
way to determine anesthesia depth, the standard was derived from measured blood
plasma concentrations of the anesthetic via a pharmacokinetic/pharmacodynamic
model for the estimated effective brain concentration of Sevoflurane. In most
patients, the highest correlation is observed for the nonlinear correlation
index~{\DAVG}.  In contrast to spectral measures, {\DAVG} is found to decrease
monotonically with increasing (estimated) depth of anesthesia, even when a
``burst-suppression'' pattern occurs in the EEG. The findings show the
potential for applications of concepts derived from the theory of nonlinear
dynamics, even if little can be assumed about the process under investigation.
\end{abstract}
\date{\today}
\section{Introduction}

Several recent publications have used analysis methods derived from the theory
of nonlinear dynamics to describe aspects of neuronal activity. Measures like
the estimated largest Lyapunov exponent (LLE) or the correlation dimension
($D_2$) were repeatedly applied~\cite{brain,Sack} to quantify brain electrical
activity, for example under the influence of epileptic processes, during
different mental states, or in different sleep stages. Attempts to interpret
the absolute numbers obtained in terms of ``system complexity'', ``attractor
dimension'', ``chaoticity'', or the like have met harsh and justified
criticism~\cite{gss}. Any such interpretation would require some narrow
assumptions (stationarity, predominant low dimensional determinism, etc.) which
are expected to be violated when biological time series are analyzed. On the
other hand, commonly applied linear or power spectral measures are based on
different, but equally narrow assumptions, including (weak) stationarity and
Gaussianity. The success of these latter methods is undisputed because of the
practical usefulness of the results. In this paper, we will try to take such a
pragmatic attitude and compare a number of indicators of human
electroencephalogram (EEG) time series with respect to their power in
monitoring depth of anesthesia.

Keeping patients at a well defined level of anesthesia is still a difficult
problem in clinical practice. If anesthesia is too deep, a decompensation of
the cardiovascular system is threatening. When anesthesia is too flat, the
patient may wake up. Depth of anesthesia is expected to be reflected in the
EEG.  Some of the underlying mechanisms are known from animal studies.  The EEG
is generated by electrical discharges (i.e. postsynaptic potentials) of neurons
located near to the surface of the brain (predominantly neurons from the
neocortex)~\cite{EEG2}. During the awake state, complex tasks can be processed
by these cells. Neuronal activity changes during anesthesia as well as during
``synchronized'' sleep (including all sleep phases except for the
rapid-eye-movement-sleep, in which most dreams are experienced). In the middle
of the brain (i.e. in the thalamus), neuronal populations start to produce an
oscillating activity that increasingly synchronizes neo-cortical
neurons. Simultaneously, the excitability of neo-cortical neurons decreases so
that the amount of interactions between them decreases as well. Finally, nearly
all neo-cortical neurons fall silent. This state will not be reached during
normal sleep, it is only observed during coma or deep anesthesia. A small
number of thalamic oscillators can even be active during this phase. Before a
continuous zero line occurs in the EEG, a so-called burst-suppression phase can
appear, where volleys from these thalamic neurons account for a cyclic
excitation of the silenced neocortex~\cite{EEG3}.

\begin{figure}
\centerline{\input{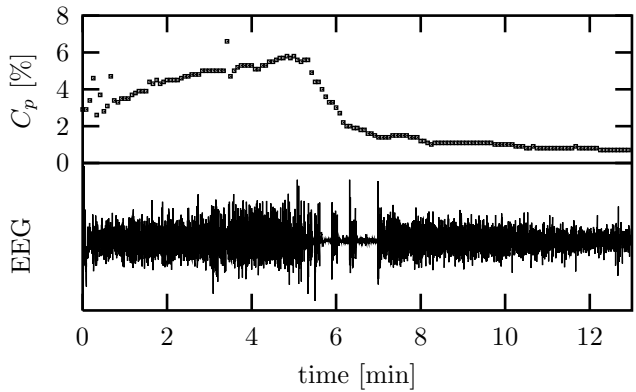}}
~\\
\caption[]{\label{fig:eeg1}
   EEG from a patient during anesthesia. The concentration of the anesthetic
   drug was increased until burst suppression first occurred (here at about
   5~min) and then reduced again. The EEG is given in uncalibrated A/D units.}
\end{figure}

The described features, except for the zero line, can be seen in
Fig.~\ref{fig:eeg1}. Note the marked hysteresis: The burst suppression
disappears only about 1~min after the plasma concentration of the anesthetic is
reduced. The scenario may be summarized by the hypothesis that with increasing
depth of anesthesia, the highly complex activities of the awake state are
increasingly superimposed and replaced by more simple oscillators.  Of course,
such a picture amounts to a gross simplification. The full mechanism will
certainly not be well described by a single parameter like ``depth of
anesthesia''.

\section{Validation of measures for depth of anesthesia}

In current clinical practice, one or a few channels of EEG are routinely
displayed during difficult anesthesias. Since the attending personnel has to
monitor several critical parameters (blood pressure, heart rate, etc.), the
vast amount of information contained in the EEG must be severely condensed in
order to be useful. Only a few numbers may be monitored at a typical
intervention time scale.  Most pragmatically, a single number should be
produced that indicates the instantaneous depth of anesthesia of the patient.

In order to derive and to validate such a measure of depth of anesthesia, the
``true'' level of anesthetic depth has to be known. One possible definition is
to take the concentration of the anesthetic drug in the brain tissue as an
indicator. This is reasonable if a monotonic relationship between the
concentration of a single drug and the depth of anesthesia can be
assumed. However, when multiple drugs are used, the drug interactions may
become unpredictable.  Unfortunately, even if a single drug is given, its brain
concentrations cannot be measured directly in human patients. In order to
circumvent this difficulty, in this study the gaseous (volatile) drug {\em
Sevoflurane} was used which can be employed as single anesthetic. In that case
the concentration of the drug in the expired air can be measured, which is
proportional to the blood-plasma concentration of Sevoflurane.  Then, by
considering the brain as a hypothetical effect compartment, the equilibration
of plasma concentration $C_p$ and the concentration in the brain $C_e$ can be
modelled in a standard way by
\begin{equation}
   \frac {d C_{e}}{d t} = {\KK\,(C_p-C_e)}  
   \label{eq:difgl}
\end{equation}
where $\KK$ is the constant of equilibration that will vary from patient to
patient. Since $\KK$ is unknown it has to be obtained by a fit of this
pharmacokinetic model to the data. Equation~(\ref{eq:difgl}) expresses the
hysteretic response noted above to a change in the amount of Sevoflurane
administered.

As a further complication, the pharmacodynamic relation between the
concentration in the brain as the effect compartment $C_e$ and the neuronal
effect $E$ is not known a priori. A reasonable assumption is that there exists
a one-to-one, monotonic relationship. In that case, and if $C_e(t)$ and $E(t)$
are sampled at discrete times sufficiently far apart to assume statistical
independence, this can be expressed by Spearman's rank
correlation. Independence is not a reasonable assumption for consecutive values
of a slowly varying quantity.  Therefore we cannot use the absolute value
of Spearman's rank correlation for a formal statistical test. However, since
the conditions are identical for all the methods we want to compare,
significant {\em differences} in the correlation may be evaluated.

The aim of our analysis here is to relate the measured $C_p$ and some estimated
value $\hat E$ of the neuronal effect $E$, assuming the kinetic model,
Eq.(\ref{eq:difgl}). Let $R_i$ be the rank of $C_e(i)$ among all available
$C_e$'s and $S_i$ the rank of $\hat E(i)$ among all the corresponding $\hat
E$'s. An estimate for $\KK$ can then be found by minimizing the sum of the
squared differences of ranks:
\begin{equation}   \label{eq:sqdr}
   \sum_{i=1}^N (R_i-S_i)^2
\,.\end{equation}
The corresponding rank correlation coefficient~\cite{stats}
\begin{equation}
   \rho=\frac{\sum_i (R_i-\overline{R})(S_i-\overline{S})}
       {\sqrt{\sum_i (R_i-\overline{R})^2} \sqrt{\sum_i (S_i-\overline{S})^2}}
   \label{eq:spearmint}
\end{equation}
quantifies the success of the fit and can thus be used as a performance index
for a method to estimate the neuronal effect.

For our study, 17 patients undergoing an elective surgery were anesthetized
with Sevoflurane. As described above, being unable to directly measure the
brain concentration $C_e$, we measured the concentration in the breathed air at
the end of expiration, which is proportional to the blood-plasma concentration
$C_p$. The neuronal effect $E$ is estimated for each patient by four
different methods based on bipolar EEG recordings from frontal scalp
regions. All EEG recordings used in this study were sampled at 256~Hz with
12~bit resolution. For the time resolved evaluation, a window length of 30~s
was used with windows taken every 5~s, that is, with 83\% overlap.
The total observation period for each patient varied between 13~min and 55~min,
median 34~min. 

\section{EEG based measures of anesthesia depth}
In this study, we have compared four methods to derive an indicator for depth
of anesthesia from EEG data. Two of these measures are based on the power
spectrum, one additionally takes the bispectrum into account, combined with a
special feature detection scheme. The fourth is a nonlinear correlation index
based on phase space reconstruction and correlation sums. Like in any other
problem of EEG 
characterisation, none of the used methods is derived from first principles nor
can any one be fully justified on physiological grounds. Rather, they aim at a
useful reduction of the vast amount of information in an EEG trace to one or a
few numbers that may be monitored with justifiable effort.

\subsection{Power spectral measures}
Apart from visual inspection, the most common method to analyse and quantify
time series data in clinical medicine --- often the only viable one with the
given data quality --- is the frequency power spectrum, usually computed using
the Fourier transform.  Spectral estimation in itself provides only a moderate
reduction of information, except in the case that dominant oscillators can be
identified by sharp spectral lines and overtones. The art of spectral analysis
is to define key parameters that turn a power spectrum into a number, or at
most a few essential quantities. For line spectra, the positions and widths of
the resonances are obvious candidates. For broad band spectra like the EEG, one
may consider the spectral content in certain frequency bands, characteristic
frequencies, etc.

Here we use two ``classical'' spectral
parameters~\cite{spectral1,spectral2,spectral3} that have been previously
applied for this purpose.  The {\em spectral edge frequency 95\%}~\cite{sef95}
indicates the maximal relevant frequencies while the {\em median
frequency}~\cite{median} gives a rough indicator over the overall typical
frequency of the recording. The physiological motivation lies in the
expectation that at deeper levels of anesthesia, high frequency components of
the surface EEG should be suppressed due to the entrainment by slow
oscillations in more central regions of the brain. Therefore, both indices are
expected to decrease with increasing depth of anesthesia. 

Both measures are based on the time windowed Fourier transform (using
5\% cosine tapering of window edges) $H(f) = \sum_{t=0}^{N-1} h(t) \exp(2\pi i
tf/N)$ of the digitally sampled EEG time series $h(t)$. Inspecting the spectral
band between $f_{\rm low} =$~0.25~Hz and $f_{\rm high} =$~30~Hz, the
accumulated periodogram
\begin{equation}
   P(f) = \sum_{i=f_{low}}^{f} |H(i)|^2
\end{equation}
is used to determine the two spectral parameters, the {\em spectral edge
frequency 95\%}, here denoted by $f_{\rm se95}$, and the {\em median
frequency} $f_{\rm median}$.  The parameter $f_{\rm se95}$ is defined as the
frequency at which $P(f_{\rm se95}) = 0.95 \, P(f_{\rm high})$ and the median
frequency $f_{\rm median}$ as the frequency at which $P(f_{\rm median}) = 0.5
\, P(f_{\rm high})$.

\subsection{Bispectral index}
A popular parameter in EEG instrumentation for anesthesia is represented by
the {\em bispectral index} ({\em bis})~\cite{rampil}. This composite parameter
uses multiple measures of the EEG power spectrum and bispectrum as well as
burst-suppression analysis and has been optimized to predict the degree of
sedation. Unlike all other EEG parameters, it has been validated in large
studies as a measure of sedation in anesthesia~\cite{sebel}. Unfortunately, the
precise algorithm is proprietary and has not yet been published. The bispectrum
is one possible natural generalisation of the Fourier power spectrum to
nonlinear signals. While the ordinary spectrum may be defined as the Fourier
transform of the two-point auto-covariance function $\langle x(t)\,x(t-\tau)
\rangle$ of a signal $x(t)$, the bispectrum is obtained by a two-dimensional
Fourier transform of the three point auto-covariance function $\langle
x(t)\,x(t-\tau_1)\,x(t-\tau_2) \rangle$. The way the bispectral index condenses
this two-variate function into a single number seems to have been established
heuristically. At least no physiological motivation is given in the literature.
In the absence of a published definition, we use the values of the bispectral
index as computed by the {\em Aspect 1000} EEG monitor and record it every 5~s.

The reported relative success~\cite{sebel} of the bispectral index suggests that
properties of the EEG beyond the power spectrum are important for 
distinguishing of different states of anesthesia. We do not want to enter
into a discussion here of what nature the extra structure is. It is very
difficult to disentangle the effects of nonlinearity and non-stationarity, and,
if nonlinearity is found, to what extent it is of dynamical nature or if it
merely reflects the complicated transduction properties of the intervening
tissue. From a physiological point of view, the situation is rather involved.
The individual neurons clearly show nonlinear behaviour in their action
potentials, but the law of large numbers suggests that this structure might be
averaged out in surface EEG traces. Non-stationarity is very prominent in
anesthesia, in particular in the burst suppression phase, where episodes of low
amplitude activity are interrupted by short outbreaks of large amplitude waves.
Spectral indicators are known to give spurious values in the presence of burst
suppression. Therefore, the bispectral index algorithm contains a pattern
recognition scheme to detect burst suppression, which is then treated
separately. 

\subsection{Nonlinear correlation index {\DAVG}}
Early claims of low dimensional strange attractors in brain signals have been
identified as spurious due to the lack of certain precautions when using
nonlinear analysis tools. (See~\cite{gss,ourbook} for further discussion and
references.) This does not exclude that tools and methodologies derived in the
context of dynamical systems, or chaos theory, may be still be useful when
employed with care. In particular, interpretations in terms of fractality,
chaoticity, complexity, and the like, have to be attempted with utmost
discretion.

In that spirit, we have adapted a prescription for an overall index of
(nonlinear) coherence that has been found powerful for anticipating epileptic
seizures from implanted electrode recordings~\cite{epileptic} as well as in
epilepsy models on a cellular level~\cite{epileptic2}.  This index, which we
will call $D^*$ contains many ingredients familiar from the
Grassberger--Procaccia algorithm for the correlation
dimension~\cite{GP,method}, whence the $D$. The $^*$ is there to remind us that
there are important differences that preclude an interpretation in terms of
fractal dimensions, number of degrees of freedom, etc.

Starting form a scalar time series $s_n, n=1,\ldots,N$, we first form
delay embedding vectors as usual: ${\bf s}_n = (s_{n-(m-1)\tau},
\ldots,s_{n-\tau},s_n)$. The correlation sum $C_m(r)$ is then
defined as usual by
\begin{equation}
   C_m(r)  = \alpha \sum_{i=1}^N\sum_{j=1}^{i-\Delta n}
   \Theta\left(r-\|{\bf s}_{i}-{\bf s}_{j}\|\right)  \label{eq:1}
\end{equation}
where $\Theta$ is the step function and $\alpha=2/(N-\Delta n)(N-\Delta n-1)$
is a normalisation constant. For the estimation of the correlation dimension,
one would take $\Delta n$ large to exclude temporal correlation and study the
slope of a double logarithmic plot of $C_m(r)$ versus $r$ and look for a {\em
scaling region}, that is, a range of values of $r$ where
\begin{equation}\label{eq:3}
   C'_m(r) = \frac{d \log(C_m(r))}{d \log(r)}
\end{equation}
is constant. For high enough embedding dimension $m$, that constant would
be an estimator of the correlation dimension $D_2$. There is ample literature
on using the correlation integral for dimension estimation~\cite{ourbook}.

Except for specific pathologies, we do not expect this procedure to arrive at a
reasonable, finite value for $D_2$ when applied to the EEG. Since we cannot
hope to estimate a proper dimension in the first place, there is no particular
reason to concentrate on geometrical correlations only, or to aim at a proper
phase space embedding. In principle, we could instead optimize parameters for
the specific purpose at hand. However, a total of 17 patients is too small a
population to be split into a training set and a test set and we are not
interested in in-sample results that may be due to over-fitting. Therefore, we
copy the parameter settings and operational procedures from previous studies on
epilepsy prediction~\cite{epileptic} without further optimization. Using unit
delay, a range of embedding dimensions $m = 10,\ldots, 25$ was studied and only
immediate temporal neighbours with $i-j<\Delta n=10$ were excluded.  Each
segment of 30~s was digitally band-pass filtered (0.25~Hz--30~Hz) before
computing~$C_m(r)$. The operational procedure of determining a ``plateau''
value $D^*$ of $C'_m(r)$ is also described in Ref.~\cite{epileptic}. It is
worth noting that this automatic procedure is designed to exclude both, large
amplitude artefacts (large $r$) and small amplitude fluctuations (small
$r$). Some conjecture on the reasons for the success of this procedure will be
offered in the discussion section below.

\section{Clinical Results}
\begin{figure}
\begin{center}
{\small}{\input{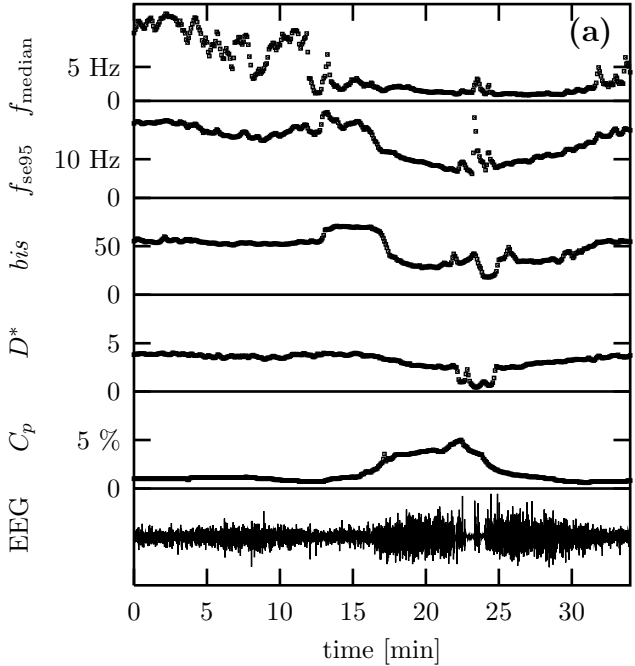}}\\[0.5cm]
{\small}{\input{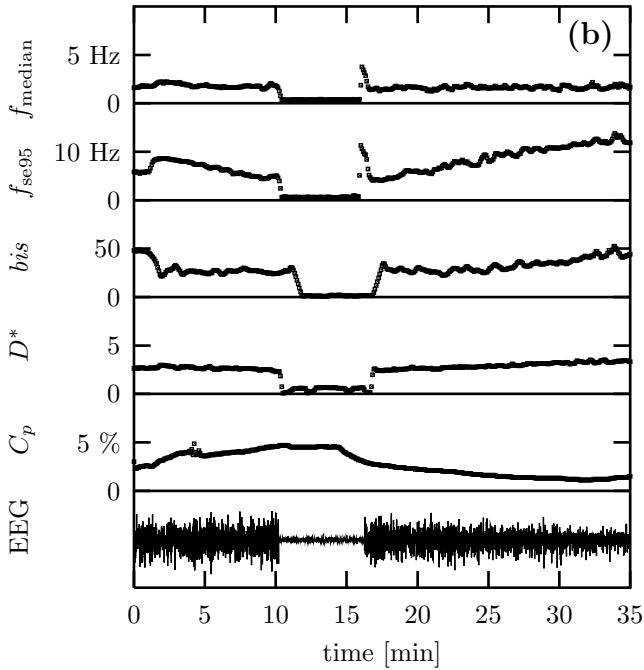}}
\end{center}
\caption[]{\label{fig:efm} Time course of two representative cases. The lowest
   trace shows the EEG and the second lowest the plasma concentration of the
   anesthetic estimated by measuring the concentration of the exhaled air.
   Depth of anesthesia is estimated every 5~s by four different methods. The
   two topmost curves denote spectral parameters, the {\em median frequency}
   $f_{\rm median}$ and the {\em spectral edge frequency 95\%}, $f_{\rm
   se95}$. The third curve shows the {\em bispectral index} ({\em bis}) as
   issued by the commercial monitoring equipment. The fourth curve gives the
   nonlinear correlation index $D^*$. See text for discussion.}
\end{figure}

In all 17 patients the following protocol was carried out.  The concentration
of Sevoflurane was at first slowly increased. As soon as a burst suppression
pattern or even a zero line occurred in the EEG, Sevoflurane concentration was
decreased again. Two typical examples of corresponding EEG time series and
estimated measures of anesthesia depth are shown in Fig.~\ref{fig:efm}. In case
B, immediately a zero line occurs in the EEG rather than a burst suppression
pattern. It is important to remember that the actual neuronal anesthetic effect
is not simply proportional to the plasma concentration $C_p$ shown in
Fig.~\ref{fig:efm} but follows hysteretically through the relaxation phenomenon
we modelled by Eq.~(\ref{eq:difgl}). Generally speaking, all four measures
studied are expected to show a negative correlation with the depth of
anesthesia.

In both cases, a slight decline of {\DAVG}, but a small increase of {\em bis}
precedes the burst suppression (resp. zero line) phase. The reaction of {\em
bis} on these abrupt changes of EEG patterns is delayed.  Depending on
frequency characteristics of the background noise, spectral measures either
decrease ({\bf A}) or increase ({\bf B}) during phases with zero-line EEG.  For
a quantitative evaluation, we need to estimate the effective brain
concentration $C_e$ of Sevoflurane. For this purpose, we assume the
pharmacokinetic model given by Eq.~(\ref{eq:difgl}) and determine an
equilibration-constant $\KK$ individually for each case by minimising the sum
of the squared differences of ranks between $C_e$ and $E$, c.f.
Eq.~(\ref{eq:sqdr}).  The values for $\KK$ that were found~\cite{foot1}
are summarized in Table~\ref{tab:k}.

\begin{table}
\begin{tabular}{lccc}
                     & median & 25\%-quartile 
                                      & 75\%-quartile  \\\hline\hline 
   $f_{\rm median}$  &  0.15  & 0.08  & 1.00 \\ 
   $f_{\rm se95}$    &  0.26  & 0.16  & 0.56 \\ 
   {\em bis}         &  0.34  & 0.22  & 0.44 \\ 
   {\DAVG}           &  0.24  & 0.18  & 0.41 \\\hline
\end{tabular}
\caption[]{\label{tab:k}
   Results of the fit for the relaxation parameter $k$.}
\end{table}

\begin{figure}
\centerline{\input{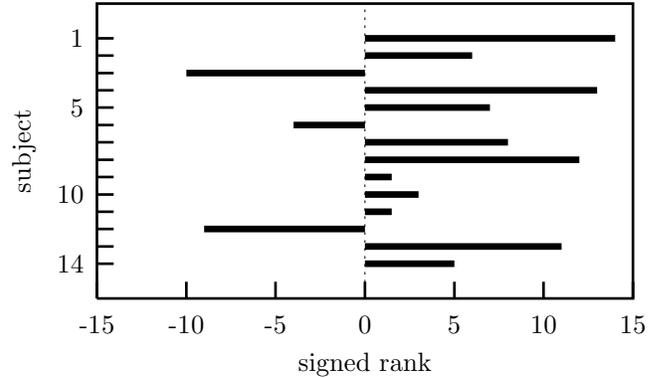}}
\caption[]{\label{fig:spear}
   Signed ranks of the differences between the correlations for {\DAVG} and
   {\em bis}. A bar to the right means that {\DAVG} correlates more strongly
   for that patient. }
\end{figure}

\begin{table}[t]
\begin{tabular}{lccc}
 & median & 25\%-quartile & 75\%-quartile \\ \hline\hline 
$f_{\rm median}$         & -0.33  & -0.06         & -0.59  \\ 
$f_{\rm se95}$           & -0.54  & -0.27         & -0.80  \\
{\em bis}            & -0.82  & -0.60         & -0.91  \\
{\DAVG}             & -0.90  & -0.81         & -0.94\\\hline
\end{tabular}
\caption[]{\label{tab:rho}
   Spearman's rank correlation $\rho$ for the best fit}
\end{table}

Compared to the variation from patient to patient, there were no dramatic
differences among the results for the different measures of anesthesia. It is
important to notice that the fit for $\KK$ is only important in order to
establish a monotonic relationship between an EEG measure and assumed depth of
anesthesia. The parameter $\KK$ describes the time course of the equilibration
of anesthetic concentration between blood plasma (assumed to be equal to the
concentration in the air at the end of expiration) and a hypothetical effect
compartment (i.e. the brain or parts of it). It will be influenced by factors
like brain perfusion, patients' age, amount of body-fat, etc.  However, since
the value of $\KK$ is no longer needed once a monotonic relationship is
assured, these influencing factors need not be under close control for this
approach.

Based on the optimal values for $\KK$, for each case Spearman's rank
correlation coefficient $\rho$ was computed (cf. Eq.~\ref{eq:spearmint}).
Table~\ref{tab:rho} shows that the highest correlation coefficients were
achieved for {\DAVG}. In order to assess the statistical significance of this
finding, we have performed a nonparametric test for the hypothesis that
{\DAVG} performs better than {\em bis}. Since the populations of values where
obtained on the same patients, we evaluated the Wilcoxon matched-pairs
signed-rank test (see Fig.~\ref{fig:spear}) and found that the correlation is
higher for {\DAVG} with $p<0.05$.

\begin{figure}[t]
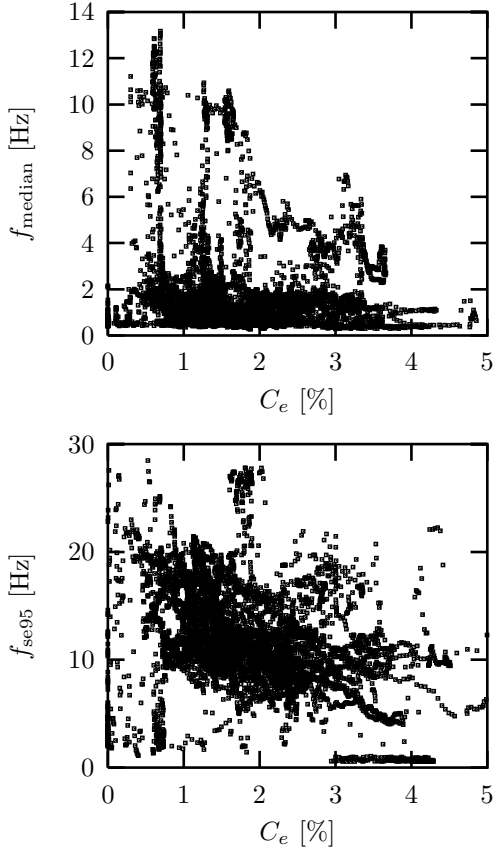

\begin{center}
\input{k_med.pst}\\
\input{k_sef.pst}
\end{center}
\caption{\label{fig:kumulat1}
   Cumulative plots of the correlation with estimated anesthesia depth
   of the two spectral indicators $f_{\rm median}$ and $f_{\rm
   se95}$, comprizing all considered cases. } 
\end{figure}

\begin{figure}[t]
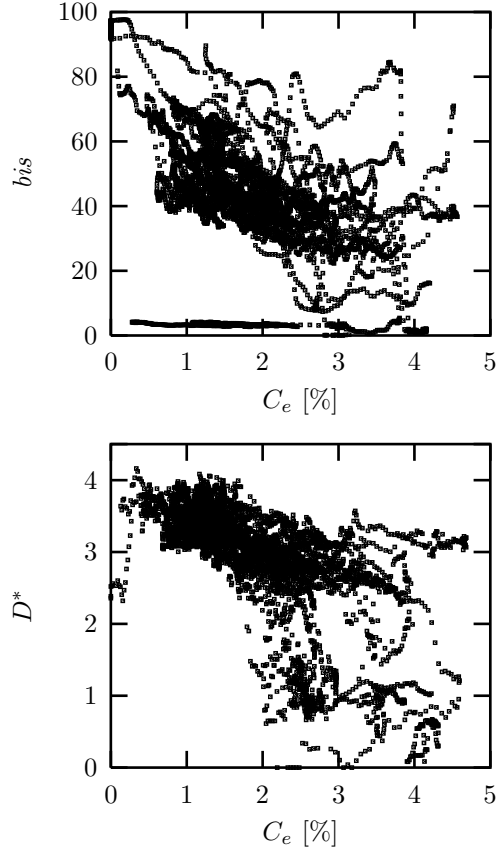

\begin{center}
\input{k_bis.pst}\\
\input{k_d.pst}
\end{center}
\caption{\label{fig:kumulat2}
   Cumulative plots of the correlation with estimated anesthesia depth
   of the two nonlinear indicators, {\em bis} and {\DAVG}, 
   comprizing all considered cases. } 
\end{figure}

For clinical applicability, it is not sufficient to have a good correlation for
each patient individually. Rather, a monotonic relationship has to hold across
patients. In order to illustrate the relevance of the relation between the
brain concentration of Sevoflurane $C_{e}$ and its estimated EEG effect $E$ for
the whole group of patients, for each investigated EEG measure a scatterplot of
$C_e$ against $E$, including results from all patients, is shown in
Figs.~\ref{fig:kumulat1} and~\ref{fig:kumulat2}. The corresponding rank
correlation coefficients $\rho$ are -0.06 for
$f_{\rm median}$, -0.33 for $f_{\rm se95}$, -0.63 for {\em bis},
and -0.72 for {\DAVG}. Due to the
serial correlations within each trial, we don't attempt a formal test of the
hypothesis that a (negative) correlation exists. Visual inspection as well as
the value for $\rho$ suggests that $f_{\rm median}$ is essentially useless as
an indicator of depth of anesthesia in the investigated group of patients Also
for the other spectral measure, $f_{\rm se95}$, the correlation is rather
unreliable. It is conceivable that in a study that excludes burst-suppression
EEG or with additional preprocessing (e.g., by introducing a burst-suppression
detector) the situation may be more favourable.  Both {\em bis} and {\DAVG}
show a certain trend to decrease with increasing depth of anesthesia. Of all
the studied measures, we find that {\DAVG} comes closest to an overall
monotonic relation to the depth of anesthesia.

%
\section{Discussion}
In a relatively small but focused clinical study, we have compared four
different methods to quantify depth of anesthesia by numerical analysis of EEG
tracings. All four methods were defined outside the study whence the results
can be considered to be out-of-sample. We have used the available data to
answer the question which of the four measures comes closest to a monotonic
relationship to the level of sedation. While for both of the two nonlinear
measures, {\em bis} and {\DAVG}, such a relationship seems to exist, the
correlation is strongest for {\DAVG}, despite the fact that {\em bis} had been
specifically designed for this purpose in previous work.

With the limited data base available, any interpretation of the findings
remains speculation. The success of {\DAVG} in previous studies on the
anticipation of epileptic seizures has been discussed in Ref.~\cite{habil},
where it was pointed out that the connection with the epileptic alterations is
lost if the temporal correlations are fully excluded from the correlation sum,
as one would have to do for a dimension estimation~\cite{ourbook}. It can be
suggested that geometrical correlations are not essential but the ability to
select structures -- temporal or geometric -- by amplitude, or length scale in
phase space, makes the correlation sum superior to other nonlinear statistics.

In earlier studies, {\DAVG} has been found helpful for the anticipation of
epileptic seizures~\cite{epileptic} from intra-cranial EEG recordings.  It has
also been shown~\cite{epileptic2} that even on a cellular level {\DAVG} could
quantify the complexity of the synaptic input of a single neuron when
inter-ictal epileptiform activity was induced in brain slices. The degree of
autonomous activity, and hence the number of degrees of freedom of a neuronal
network, is thought to decrease during the development of epileptiform
activity. Indeed, a reduction of {\DAVG} extracted from intracellular
recordings of single neurons was observed prior to visually detectable
alterations~\cite{epileptic2}.

In the common scenario of anesthesia, the number of degrees of freedom of human
brain activity is lowered --- as compared to the awake state --- when nearly
all neo-cortical neurons fall silent and only a central (thalamic) oscillator
is able to evoke a cyclic excitation of the silenced
neocortex~\cite{EEG3}. However, the number of degrees of freedom may still be
huge and is probably not accessible from an EEG time series~\cite{ruelle}. It
is even the more remarkable that {\DAVG} correctly ranks the qualitatively
different EEG pattern of burst suppression as corresponding to the deepest
level of anesthesia.  At the onset of burst suppression, the EEG signature
changes discontinuously with the estimated drug concentration, whence the
relatively sharp transition in {\DAVG} between the preceding deep sleep phases
and burst suppression in Fig.~\ref{fig:kumulat2}. Still, from a biological
point of view it remains unclear which electrophysiological properties
contribute to the monotonic decline of {\DAVG} under anesthesia --- or
epileptic activity.

In summary, extraction of {\DAVG} seems to be able to improve the
quantification of depth of anesthesia from brain electrical activity, at least
when Sevoflurane is used as anesthetic drug. For clinical applicability, other
anesthetics have to be investigated. The size of this study is too small to
admit any optimisation of the algorithm. It is conceivable, however, that the
clinical reliability can be enhanced, and the computational burden relieved, by
a more thorough understanding of which structures are really picked up by the
algorithm.

\end{document}